\documentclass[fleqn,12pt,twoside]{article}
\usepackage{espcrc1}

\usepackage{graphicx}
\usepackage[figuresright]{rotating}

\newcommand{\AmS}{{\protect\the\textfont2
  A\kern-.1667em\lower.5ex\hbox{M}\kern-.125emS}}

\def\lsi{\raise0.3ex\hbox{$<$\kern-0.75em\raise-1.1ex\hbox{$\sim$}}}
\def\gsi{\raise0.3ex\hbox{$>$\kern-0.75em\raise-1.1ex\hbox{$\sim$}}}

\newcommand{\gsim}{\mathop{\gsi}}

\title{Simulating chiral quarks in the $\epsilon$-regime of QCD}

\author{W. Bietenholz
\address{Institut f\"{u}r Physik, Humboldt Universit\"{a}t zu Berlin,
Newtonstr. 15, D-12489 Berlin, Germany \\
$^{{\rm b}}$ NIC/DESY Zeuthen, Platanenallee 6, D-15738 Zeuthen, Germany}
, T. Chiarappa $^{{\rm b}}$, K. Jansen $^{{\rm b}}$,
K.-I. Nagai $^{{\rm b}}$ and S. Shcheredin $^{{\rm a}}$  
}

\begin{document}

\maketitle

\begin{abstract}
We present simulation results for lattice QCD with chiral fermions
in small volumes, where the $\epsilon$-expansion of chiral perturbation
theory ($\chi$PT) applies. Our data for the low lying Dirac eigenvalues,
as well as mesonic correlation functions, are in agreement with
analytical predictions. This allows us to extract values for the 
leading Low Energy Constants $F_{\pi}$ and $\Sigma$.
\end{abstract}

\vspace*{-3mm}
\section{The $\epsilon$-regime of QCD}
\vspace*{-2mm}

For QCD at low energy and zero quark masses one assumes the 
chiral symmetry to break spontaneously, $SU(N_{f})_{L} \otimes
SU(N_{f})_{R} \to SU(N_{f})_{L+R}$. This yields $N_{f}^{2}-1$ 
Nambu-Goldstone bosons, which pick up a small mass when we proceed
to light quark masses $m_{q}$ (we assume the same $m_q$ for all flavours
involved). These bosons are identified with 
the light mesons, in particular the pions (and often also the kaons
and $\eta$). 

$\chi$PT is a powerful description for them by means of an effective 
chiral Lagrangian, 
\begin{displaymath}
{\cal L}_{\rm eff} [U] = \frac{F^{2}_{\pi}}{4} \, {\rm Tr}
[\partial_{\mu} U \partial_{\mu} U^{\dagger} ] - \frac{m_{q}}{2} \,
\Sigma \, {\rm Tr} [ U + U^{\dagger}] + \dots \ ,
\end{displaymath}
which deals with fields $U$ in the coset space
$SU(N_{f})$ of the chiral symmetry breaking.
One writes down the terms which fulfil the symmetry requirements and
classifies them according to low energy counting rules. The coefficients 
to these terms are the Low Energy Constants (LEC), 
hence $F_{\pi}$ and $\Sigma$ are the leading LEC.

One usually considers this effective theory in an infinite or very large
volume, which allows for an expansion in the meson momenta
($p$-expansion). However, one can also consider the opposite situation ---
a small volume with linear size $L \ll 1 / m_{\pi}$ --- which is called the
{\em $\epsilon$-regime}. Here the $p$-expansion fails due to the importance 
of the zero modes. Fortunately the path integral for the latter
can be performed with collective variables \cite{GasLeu}. 
Then an expansion in the low (non-zero) momenta and
the meson mass is possible ($\epsilon$-expansion).

It is in fact motivated to study the $\epsilon$-regime, although
it is not a physical situation to squeeze pions into such a tiny box.
The point is that even in a small volume values for the LEC in infinite
volume can be determined --- exactly by analysing the finite size effects.

The LEC enter the effective Lagrangian as free parameters, hence
they can only be determined from QCD as the fundamental theory (or
from experiment). This is a goal for lattice simulations, which are
clearly more convenient in small volumes.

However, simulations in the $\epsilon$-regime face technical difficulties,
which could only be overcome in the recent years. One has to keep track
of the chiral symmetry and deal with very light pions. Moreover, we need
a clean definition for the topology, since expectation values should
often be measured in distinct topological sectors \cite{LeuSmi}.
These properties are provided by Ginsparg-Wilson fermions,
which have a lattice modified, exact chiral symmetry \cite{ML} and exact
zero modes, so that the topological charge can be defined as the index
$\nu$ (i.e.\ the difference between the number of zero modes with
positive and negative chirality) \cite{Has}. The simplest realization
of a Ginsparg-Wilson fermion uses the overlap Dirac operator \cite{Neu}.

Although algorithmic tools for the $\epsilon$-regime have been worked out 
\cite{algo}, such simulations are still computationally expensive: 
for the time being one has to use the quenched approximation --- which 
leads to logarithmic finite size effects on the LEC \cite{quench}.

\vspace*{-2mm}
\section{The Dirac spectrum}
\vspace*{-3mm}

Chiral Random Matrix Theory (RMT) simplifies QCD to a Gaussian distribution of
fermion matrix elements, and still manages to capture some aspects of
QCD. One of its numerous assumptions is that the energy should be
well below the Thouless energy 
$E_{\rm Thouless} = F^{2}_{\pi} / (\sqrt{V} \Sigma )$.

\begin{figure}[htb]
\includegraphics[angle=270,scale=0.45]{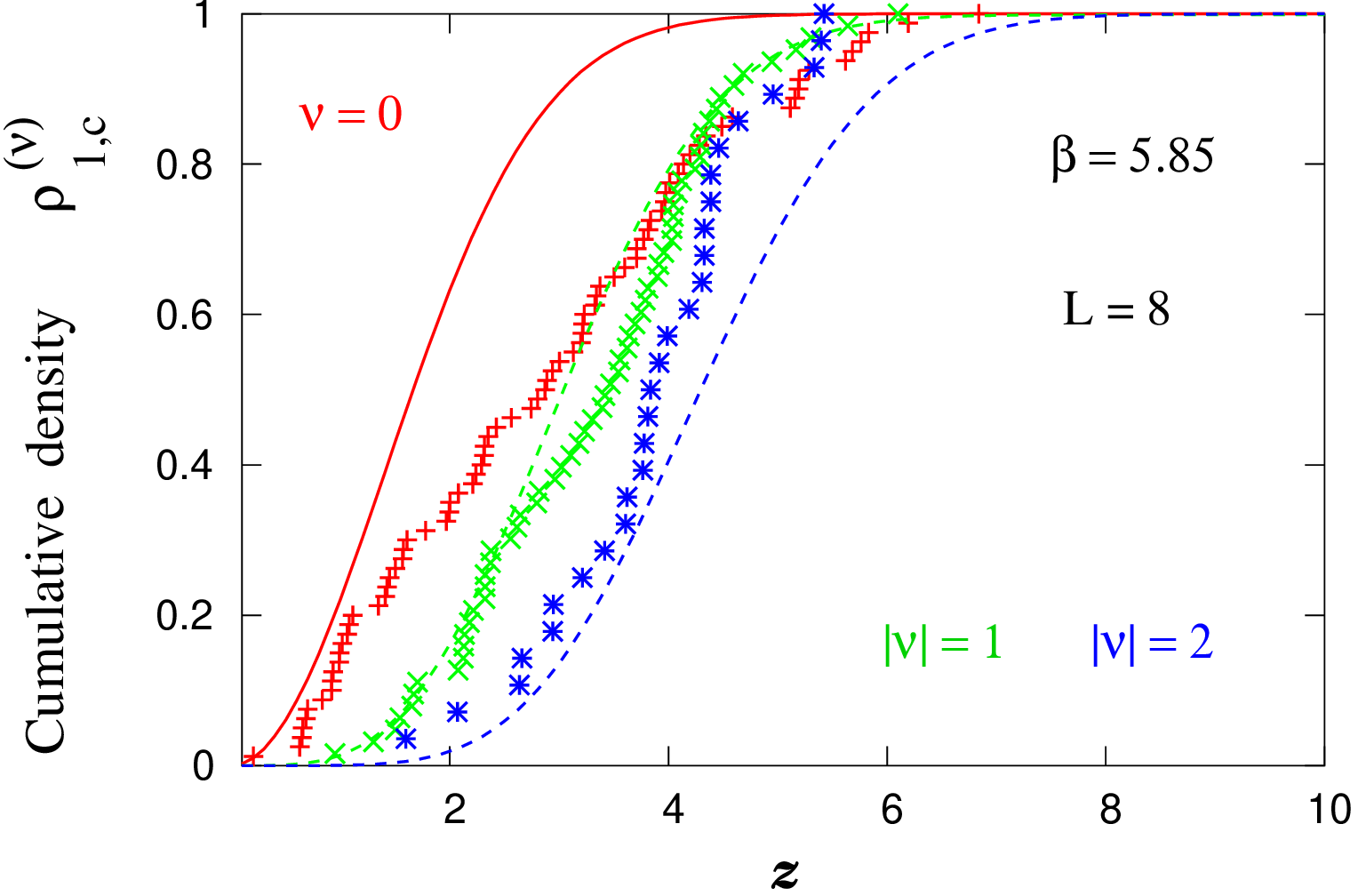} 
%
%
\includegraphics[angle=270,scale=0.45]{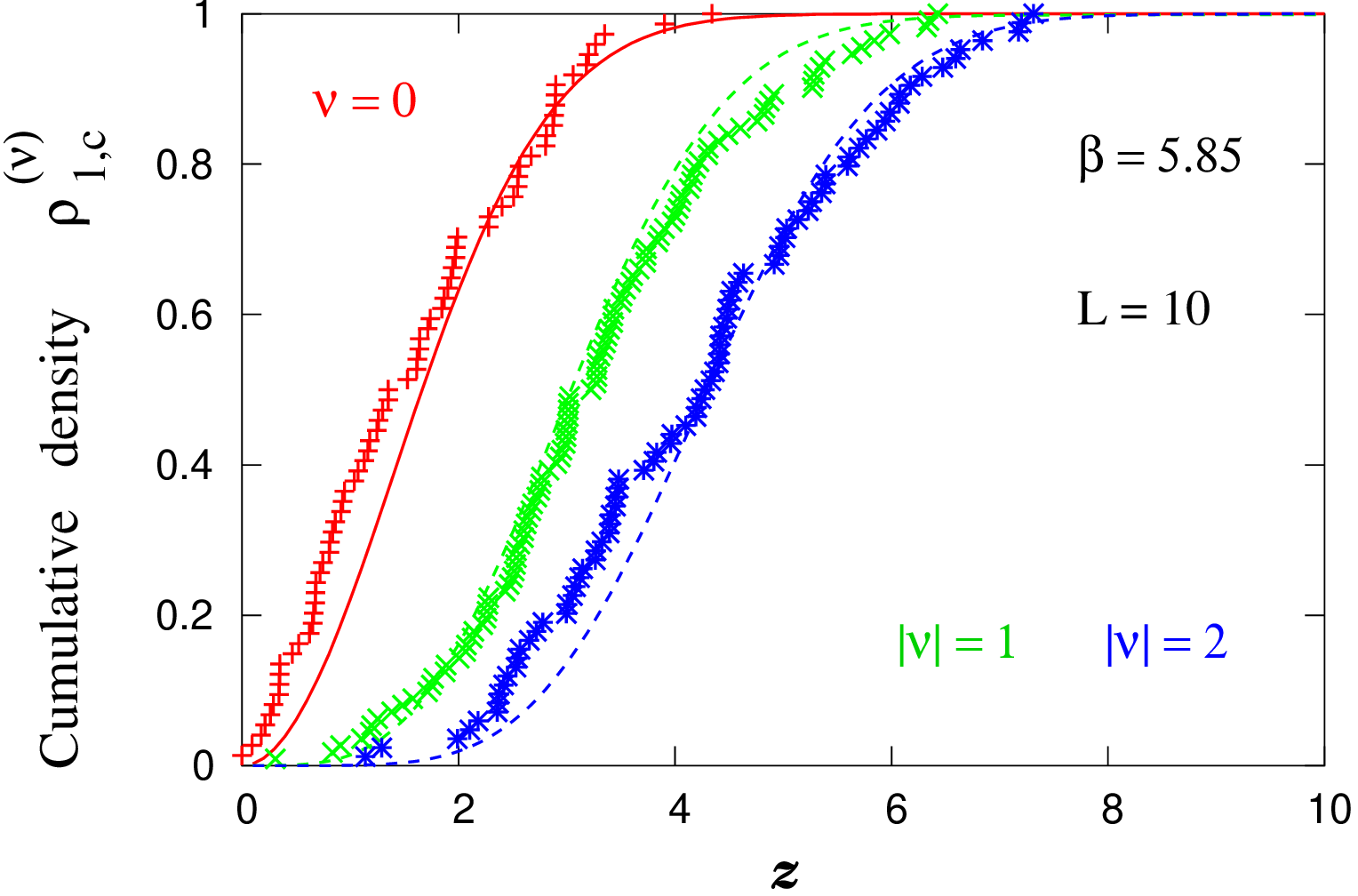} 
\vspace*{-8mm}
\caption{\it{The cumulative density of the first non-zero eigenvalue
in the sectors
$\vert \nu \vert =0,1$ and $2$. The comparison of our lattice data
to the chiral RMT predictions (curves) fails in $V = (0.98~{\rm fm})^{4}$
(on the left), but it is successful in $V = (1.23~{\rm fm})^{4}$ 
(on the right). In the latter case, the fit yields $\Sigma = 
(253~{\rm MeV})^{3}$.}}
\label{cumdense}
\vspace{-8mm}
\end{figure}

In particular, Ref.\ \cite{RMT} presents explicit predictions for the 
microscopic spectral density
\begin{displaymath}
\rho_{s}(z) = \left. \frac{1}{\Sigma V} \rho \Big( \frac{z}{\Sigma V} \Big) 
\right\vert_{V \to \infty} \!\!
= \sum_{\nu = 0}^{\infty} \rho_{s}^{(|\nu |)}(z) , \
z = \lambda \Sigma V, \ \lambda \ : \, 
{\rm eigenvalue~of~the~Dirac~operator.}
\end{displaymath}
The density in each topological sector $| \nu |$ is further decomposed
into densities of the leading individual eigenvalues, 
$\rho_{s}^{(|\nu |)}(z) = \sum_{n = 1,2,3 \dots}\rho_{n}^{(|\nu |)}(z)$.
Using the overlap operator at $m_q = 0$ we have measured 
$\rho_{1}^{(|\nu |)}(z)$ at 
$|\nu | = 0,\ 1$ and $2$ in different volumes \cite{EV1}. In Figure
\ref{cumdense} we compare our data for the cumulative density 
with the RMT predictions. The comparison fails in the small volume
$V \simeq (0.98 ~ {\rm fm})^{4}$, but it works well in
$V \simeq (1.23 ~ {\rm fm})^{4}$. In this fit, $\Sigma$ appears as the only 
free parameter. The successful fit in $V \simeq (1.23 ~ {\rm fm})^{4}$
implies a value of $\Sigma \simeq (253~{\rm MeV})^{3}$, which agrees well
with the literature.

The microscopic RMT predictions work in such volumes up to the second
eigenvalue, then the spectral density starts to grow cubically
(bulk behaviour). This transition can be viewed as an effective 
bound for the chiral RMT, thus taking the r\^{o}le of 
$E_{\rm Thouless}$. Our observations were confirmed by
subsequent studies, which 
included larger volumes \cite{EV2}.

\vspace*{-3mm}
\section{Axial vector correlation functions}
\vspace*{-2mm}

Since QCD simulations with chiral fermions are restricted to the
quenched approximation at present, also quenched $\chi$PT was worked
out \cite{qXPT} as a basis of predictions that the lattice data
can be confronted with. Regarding the mesonic correlation functions 
in leading order, the vector correlator vanishes (actually to all orders), and
the scalar and pseudoscalar correlators have the drawback that they
involve additional LEC, 
which are specific to quenching. Hence we considered the correlator of
the axial current $A_{\mu}(t) = \sum_{\vec x} \bar \Psi (t, \vec x)
\gamma_{5} \gamma_{\mu} \Psi (t, \vec x )$, where the leading order only
involves $F_{\pi}$ and $\Sigma$. The $\epsilon$-regime prediction
for $\langle A_{0}(t) A_{0}(0) \rangle _{| \nu |}$ in a volume
$L^{3} \times T$ is a parabola in $t$ with a minimum at $T/2$,
and with an additive constant $F_{\pi}^{2}/T$ (this is in contrast to
the usual {\tt cosh} behaviour in large volumes).
The curvature
in the minimum of the parabola is $\propto \Sigma$.
We now simulated overlap fermions at $m_q = 21.3 ~{\rm MeV}$
\cite{AA}.
A first experiment with $L \simeq 0.98~{\rm fm}$ failed to reproduce
the prediction. This is exactly consistent with the observation
for the Dirac eigenvalues described before: in both cases
we badly need $L \gsim 1.1 ~{\rm fm}$.

We also observed that the Monte Carlo history at $\nu =0$ is plagued 
by spikes, so that a huge statistics is needed. These spikes occur 
at the configurations which have a very small (non-zero) Dirac eigenvalue.
As Figure \ref{cumdense} shows, this probability is strongly suppressed
for $|\nu | > 0$, and indeed these histories are smoother.

\begin{figure}[htb]
\vspace*{-8mm}
\begin{center}
\includegraphics[angle=270,scale=0.42]{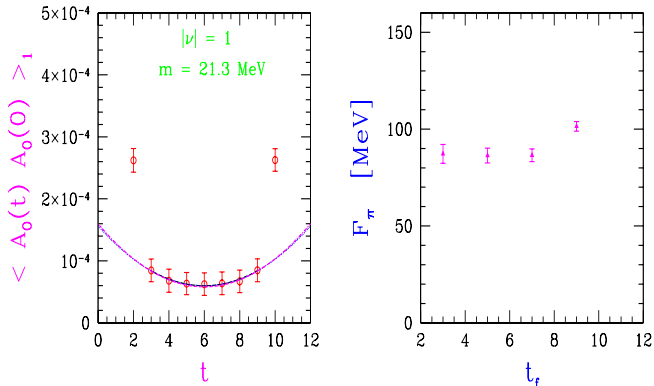}
\end{center}
\vspace*{-14mm}
\caption{\it{The axial correlation function (on the left) and the
values of $F_{\pi}$ obtained from fits over $t_{f}$ points around the 
centre $t=6$ (on the right).}}
\label{AAfig}
\end{figure}

So we measured in a volume $V = (1.12~{\rm fm})^{4}$ at $| \nu | =1$.
Figure \ref{AAfig} (on the left) shows that we found a nice 
agreement with the predicted parabola, which is plotted for 
$\Sigma = 0$ and for $\Sigma = (250 ~{\rm MeV})^{3}$. 
Since these curves can hardly be distinguished, we see that
the sensitivity of the curvature is by far too small for a
determination of $\Sigma$ in this way. On the other hand,
this result provides a stable value of $F_{\pi} = (86. 7 \pm 4.0)
~{\rm MeV}$. The plot on the right-hand-side of Figure \ref{AAfig} 
shows the value we obtained as a function of the fitting range
around the minimum. The result that we obtained for $| \nu | =2$
(with lower statistics) agrees within the errors.
The renormalisation due to the factor $Z_{A}$ drives
our result for $F_{\pi}$ up to a value around $130 ~{\rm MeV}$,
but it is well-known that quenched results for the pion decay constant
tend to be significantly above the experimental value of $\approx 93 
~{\rm MeV}$ resp.\ $\approx 86 ~{\rm MeV}$ in the chiral limit \cite{CD}.

\vspace*{-3mm}
\section{Conclusions}
\vspace*{-2mm}

We have performed two pilot studies about the feasibility to determine
LEC of $\chi$PT from simulations with chiral lattice fermions
in the $\epsilon$-regime. Our results show that such lattice data
{\em can} be fitted to the prediction by chiral RMT and quenched $\chi$PT.
However, we also found crucial conditions for this to work:
the spatial box length must exceed a lower bound, 
$L \gsim 1.1 ~ {\rm fm}$. Moreover, correlators should not be
measured in the topologically trivial sector.

$F_{\pi}$ can be extracted from the axial vector
correlator, and $\Sigma$ from the densities of the lowest eigenvalues
of the Dirac operator. Thus our pilot studies confirm that simulations
in the $\epsilon$-regime do have the potential to evaluate LEC
from first principles. One could also deal solely with the
zero-mode contributions to the mesonic correlation functions
\cite{zeromodes}.

Currently a non-standard lattice gauge action is being tested,
which preserves the topological charge over long intervals of the
Monte Carlo history (far beyond the autocorrelation time of other
observables) \cite{topogauge}.
This would allow us to collect many configurations for measurements
in a specific topological sector, which is tedious with the standard 
gauge action. On the fermionic side, we are now including an improved
overlap operator \cite{ovimp} in our investigation of the $\epsilon$-regime.
At last we add that we also have results in the $p$-regime \cite{preg},
for instance for the direct measurement of $F_{\pi}$.

{\it This work was supported by the Deutsche Forschungsgemeinschaft 
through SFB/TR9-03. 
Most computations were performed at the ZIB/HLRN in Berlin.
}

\end{document}